\documentclass[useAMS,usenatbib]{mn2e}
\usepackage{fancyhdr}
\usepackage{graphicx}
\usepackage{times}
\usepackage{url}
\newcommand{\HI}{\hbox{\rmfamily H\,{\textsc i}}}
\newcommand{\HIsub}{\hbox{{\scriptsize H}\,{\tiny I}}}
\newcommand{\MHI}{\hbox{$M_{\HIsub}$}}
\newcommand{\rhoHI}{\hbox{$\rho_{\HIsub}$}}
\newcommand{\OHI}{\hbox{$\Omega_{\HIsub}$}}
\newcommand{\msun}{\hbox{${\rm M}_{\odot}$}}
\newcommand{\lsun}{\hbox{${\rm L}_{\odot}$}}
\newcommand{\Ha}{\hbox{\rmfamily H\,$\alpha$}}
\newcommand{\kms}{\hbox{km\,s$^{-1}$}}

\title[${\HI}$ observation of the COSMOS field at $z \sim$~0.37]{GMRT observation of neutral atomic hydrogen gas in the COSMOS field at $z \sim$~0.37}

\author[J. Rhee et al.]{Jonghwan Rhee$^{1,2,4}$\thanks{E-mail: jonghwan.rhee@icrar.org, now at ICRAR}, Philip Lah$^{1,3}$,  Jayaram N. Chengalur$^{3}$, Frank H. Briggs$^{1,4}$, \newauthor and Matthew Colless$^{1,4}$\\ 
$^{1}$Research School of Astronomy and Astrophysics, Australian National University, Canberra, ACT 2611, Australia\\
$^{2}$International Centre for Radio Astronomy Research (ICRAR), University of Western Australia, Crawley, WA 6009, Australia\\
$^{3}$National Centre for Radio Astrophysics, Tata Institute for Fundamental Research, Pune 411 007, India\\
$^{4}$ARC Centre of Excellence for All-sky Astrophysics (CAASTRO)}

\begin{document}

\date{Accepted 2016 May 06. Received 2016 May 05; in original form 2015 October 08}

\pagerange{\pageref{firstpage}--\pageref{lastpage}} \pubyear{2016}

\maketitle

\label{firstpage}

\begin{abstract}
We present the results of {\HI} spectral stacking analysis of  Giant Metrewave Radio Telescope (GMRT) observations targeting the COSMOS field. The GMRT data cube contains 474 field galaxies with redshifts known from the zCOSMOS-bright 10k catalogue. Spectra for the galaxies are co-added and the stacked spectrum allows us to make a $\sim 3\sigma$ measurement of the average {\HI} mass. Using this average {\HI} mass along with the integral optical $B$-band luminosity of the galaxies and the luminosity density of the COSMOS field, a volume normalisation is applied to obtain the cosmic {\HI} mass density ({\OHI}). We find a cosmic {\HI} mass density of {\OHI}~$=$~(0.42~$\pm$~0.16)~$\times$~10$^{-3}$ at $z \sim 0.37$, which is the highest-redshift measurement of {\OHI} ever made using  {\HI} spectral stacking. The value we obtained for {\OHI} at $z \sim 0.37$ is consistent with that measured from large blind 21-cm surveys at $z = 0$ as well as measurements from other {\HI} stacking experiments at lower redshifts. Our measurement in conjunction with earlier measurements indicates that there has been no significant evolution of {\HI} gas abundance over the last 4 Gyr. A weighted mean of {\OHI} from all 21-cm measurements at redshifts $z \la 0.4$  gives {\OHI}~$=$~(0.35~$\pm$~0.01)~$\times$~10$^{-3}$. The {\OHI} measured (from {\HI} 21-cm emission measurements) at $z \la 0.4$ is however approximately half that measured from Damped Lyman-$\alpha$ Absorption (DLA) systems at $z \ga 2$. Deeper surveys with existing and upcoming instruments will be critical to understand the evolution of {\OHI} in the redshift range intermediate between $z \sim 0.4$ and the range probed by DLA observations.
\end{abstract}


\section{INTRODUCTION}

Our knowledge of the evolution of galaxies has been almost exclusively acquired by large-scale surveys using ground and space-based telescopes
at optical and adjacent wavelengths \citep[e.g.,][]{York:2000,Colless:2001,Martin:2005,Skrutskie:2006}. The information obtained from these surveys is mainly valuable for tracing the evolution of the stellar content of galaxies. In contrast, little is known about the evolution of the (cold) gaseous component in galaxies. In particular, our knowledge of the neutral atomic hydrogen gas ({\HI}) content (i.e. the raw material for star formation) of individual galaxies is essentially limited to the very low redshift universe \citep{Meyer:2004,Haynes:2011}.

Understanding the evolution of the atomic gas content of galaxies remains one of the key challenges in the study of galaxy evolution. The sensitivity of the current generation of radio telescopes is insufficient to detect {\HI} from individual galaxies at cosmologically significant redshifts in reasonable integration times. Indeed, the detection of {\HI} from individual galaxies at high redshifts was one of the original motivations, and remains one of the key science drivers for the proposed Square Kilometre Array \citep[SKA, e.g.,][]{Blyth:2015,Staveley-Smith:2015,Santos:2015}. It is also one of the key programs for several of the upcoming SKA pathfinder telescopes, for instance the Widefield ASKAP L-band Legacy All-sky Blind surveY \citep[WALLABY,][]{Koribalski:2009}, the Deep Investigation of Neutral Gas Origins \citep[DINGO,][]{Meyer:2009}, and Looking At the Distant
Universe with the MeerKAT Array \citep[LADUMA,][]{Holwerda:2012}. In combination with data at other wavelengths, this knowledge of {\HI} gas
obtained from the SKA and its pathfinders will allow us to fully understand the complex processes that govern galaxy evolution \citep{Meyer:2015}.

Although it is challenging to detect {\HI} in individual galaxies at $z \ga 0.2$ with the current generation of radio telescopes, it is possible to make measurements of the average {\HI} content of a sample of galaxies. The volume of space probed in a single pointing and correlator setting typically contains many hundreds of galaxies. If the positions and redshifts of all of these galaxies are known, one can stack their spectra to determine their average {\HI} content \citep{Chengalur:2001,Zwaan:2001}. This spectral stacking technique has been applied to interferometers such as the Giant Metrewave Radio Telescope (GMRT) and the Westerbork Radio Telescope (WSRT) as well as single dishes such as the Parkes to measure the {\HI} content of galaxies at redshifts $z \la 0.4$, resulting in determination of the evolution of the gas content in galaxies (i.e. {\OHI}) out to redshifts
$z \sim 0.2$ \citep{Lah:2007, Lah:2009,Delhaize:2013,Rhee:2013}.

Here we apply the technique to GMRT observations of the COSMOS field to determine {\OHI} at a redshift of $z\sim 0.37$. The GMRT observation of
the COSMOS field plays an important role as a precursor of future {\HI} deep surveys using SKA pathfinders. To check the viability of the surveys and develop suitable observing modes and strategies, a smaller-scale {\HI} survey, as presented here, is a good testbed. Both expected and unexpected issues relevant to wide and deep {\HI} surveys can be explored through such a pilot study. The redshift studied ($z \sim$~0.37 ) allows us to understand how the {\HI} gas in galaxies evolves out to $z \sim$~0.4, and perhaps beyond.

This paper is structured as follows: In Section~2 we detail the optical data of the COSMOS field. Section~3 describes the GMRT observations, data quality and data reduction procedures. We explain the stacking analysis used to measure the average {\HI} gas mass at $z \sim$~0.37 in Section~4. The main results are presented in Section~5. In Section~6, we discuss the implications of our observations for the cosmic evolution of {\OHI}. We present the summary and conclusions in Section~7. Throughout the paper we adopt the concordance cosmological parameters: $\Omega_{\Lambda}=$~0.7, $\Omega_{M}=$~0.3 and $H_{0}=$~70~km~s$^{-1}$~Mpc$^{-1}$.

\section{Data}

\subsection{Target Field Selection}
Our radio observations are centred on the COSMOS field. This field has a wealth of multi-wavelength data ranging from X-ray to radio \citep{Scoville:2007}. Spectroscopic redshifts are also available for a very large number of galaxies in the COSMOS field \citep[zCOSMOS,][]{Lilly:2007}. This makes it an excellent target for {\HI} studies. Indeed it has also been selected as the target for a very deep JVLA\footnote{The Karl G. Jansky Very Large Array} {\HI} survey \citep[CHILES,][]{Fernandez:2013}. 

\begin{table}
 \centering
 \caption[COSMOS photometry band filters and their zero-point offsets]{Photometric band filters and their zero-point offsets in the COSMOS photometry catalogue taken from \citet{Ilbert:2009}. All magnitudes of our sample in these photometric bands were used as inputs for the SED fitting procedure with {\sc le phare}.} 
 \begin{tabular}{@{}lcccr}
  \hline
  Filter & Telescope  & $\lambda_{\rm eff}$ & FWHM & offset \\
           &                   & ($\AA$)                 & ($\AA$) &          \\  
 \hline
 FUV       & $GALEX$   & 1551.3       & 230.8  &  0.314  \\
 NUV      & $GALEX$   & 2306.5       & 789.1  & -0.022  \\
 $u^{*}$  & CFHT         & 3911.0       & 538.0  &  0.054  \\
 $B_{J}$   & Subaru       & 4439.6       & 806.7  & -0.242  \\
 $V_{J}$   & Subaru       & 5448.9       & 934.8  & -0.094  \\ 
 $g^{+}$ & Subaru       & 4728.3       & 1162.9 &  0.024  \\
 $r^{+}$  & Subaru       & 6231.8       & 1348.8 &  0.003  \\
 $i^{+}$  & Subaru       & 7629.1       & 1489.4 &  0.019  \\
 $i^{*}$   & CFHT         & 7628.9       & 1460.0 & -0.007  \\
 $z^{+}$ & Subaru       & 9021.6       &  955.3  & -0.037  \\
 $J$       & UKIRT         & 12444.1     & 1558.0 &  0.124  \\
 $K_{s}$  & KPNO/CTIO & 21434.8    & 3115.0 &  0.022  \\
 $K$      & CFHT          & 21480.2     & 3250.0 & -0.051  \\
 $IA$427 & Subaru     & 4256.3       & 206.5   &   0.037  \\
 $IA$464 & Subaru     & 4633.3       & 218.0   &   0.013  \\  
 $IA$484 & Subaru     & 4845.9       & 228.5   &   0.000  \\
 $IA$505 & Subaru     & 5060.7       & 230.5   & -0.002  \\
 $IA$527 & Subaru     & 5258.9       & 242.0   &   0.026  \\ 
 $IA$574 & Subaru     & 5762.1       & 271.5   &   0.078  \\ 
 $IA$624 & Subaru     & 6230.0       & 300.5   &   0.002  \\
 $IA$679 & Subaru     & 6778.8       & 336.0   & -0.181  \\
 $IA$709 & Subaru     & 7070.7       & 315.5   & -0.024  \\
 $IA$738 & Subaru     & 7358.7       & 323.5   &   0.017  \\
 $IA$767 & Subaru     & 7681.2       & 364.0   &   0.041  \\
 $IA$827 & Subaru     & 8240.9       & 343.5   & -0.019  \\
 $NB$711 & Subaru    & 7119.6       &  72.5    &   0.014  \\
 $NB$816 & Subaru    & 8149.0       & 119.5   &   0.068  \\
 \hline
 \end{tabular}
 \label{tab:cosmos_cat}
\end{table}

\begin{figure*}
\centering
\includegraphics[width=150mm]{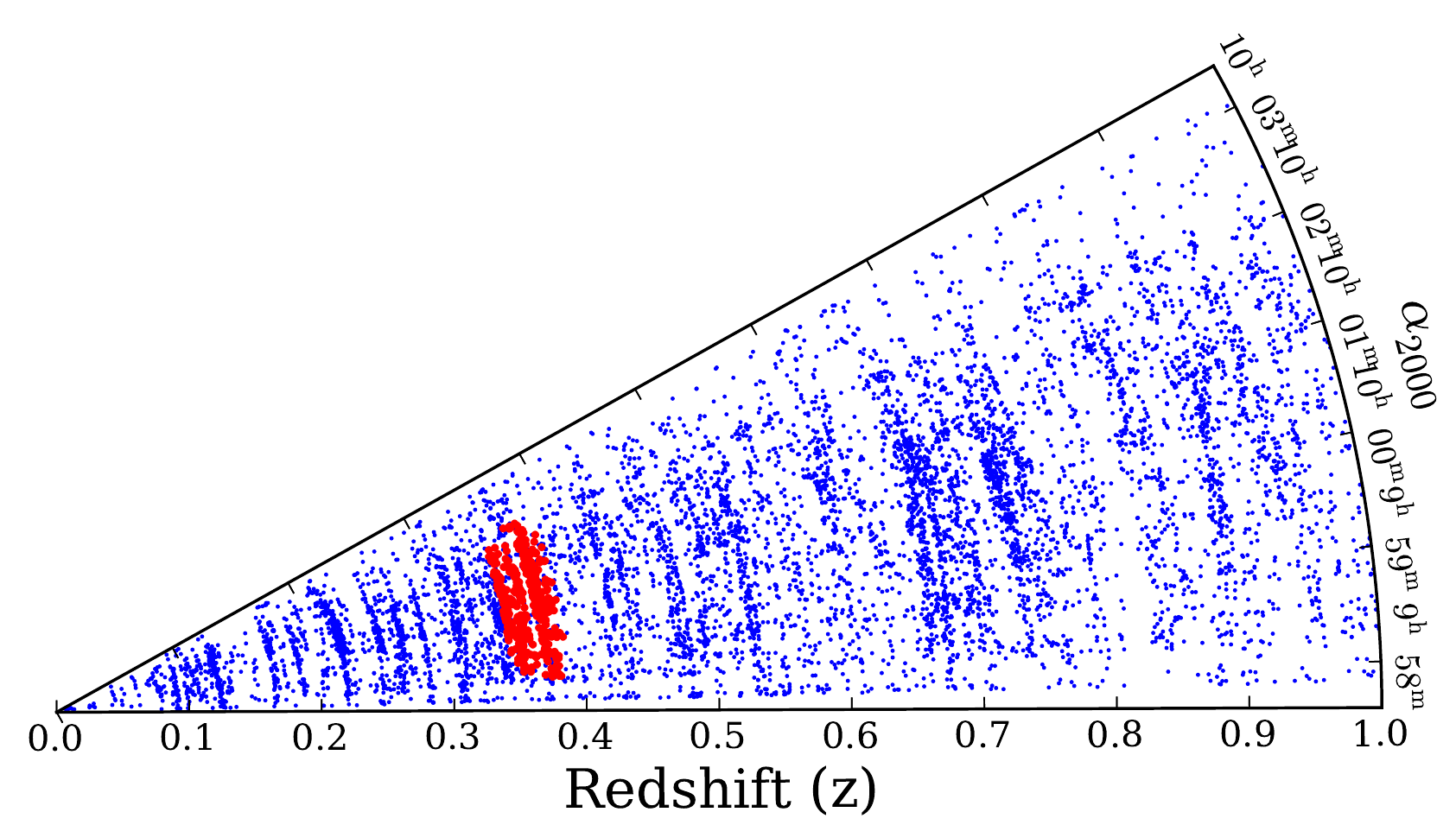} 
\caption[Redshift cone diagram of zCOSMOS field]{Redshift cone diagram of zCOSMOS 10k-bright sample to $z \sim$~1. Red points denote the 506 galaxies covered by GMRT beam and frequency at $z \sim$~0.37.}
\label{fig:z_diagram}
\end{figure*}

\subsection{Photometric Data}
Optical positions and redshifts are crucial inputs required for {\HI} stacking analysis. Multi-band photometric data is also very important, because it allows one to determine the cosmic {\HI} density as well as the dependence of {\HI} gas content on other galaxy properties such as the morphological or spectral type, etc. Photometry of the galaxies in the COSMOS field is available over a very large range of wavelengths: X-ray with {\it XMM-Newton} \citep{Hasinger:2007}, UV with {\it GALEX} \citep{Zamojski:2007}, optical/NIR with CFHT, CTIO, KPNO, Subaru \citep{Taniguchi:2007} and {\it Hubble Space Telescope} \citep[{\it HST},][]{Koekemoer:2007}, mid-infrared with the {\it Spitzer} space telescope \citep{Sanders:2007}, mm/sub-millimetre with the Caltech Submillimeter Observatory (CSO) and Institut de Radioastronomie Millim\'{e}trique (IRAM) telescope \citep{Bertoldi:2007} and radio continuum with the Very Large Array \citep[VLA,][]{Schinnerer:2004, Schinnerer:2007}. 

We  cross-matched the zCOSMOS redshift catalogue with the publicly released photometric catalogue of the COSMOS field from the {\it IRSA} website\footnote{NASA/IPAC Infrared Science Archive, {\url {http://irsa.ipac.caltech.edu/data/COSMOS/}}}. This provides  observed magnitudes in 27 photometric bands covered by CFHT, Subaru, KPNO/CTIO, and {\it GALEX}. This catalogue is an update on the previous optical/NIR catalogue by \citet{Capak:2007}. The updated photometric catalogue was compiled using the same point-spread function (PSF)   from $u^*$ to the $K$ band. The photometry was derived over the same aperture of 3\arcsec diameter centred on the position of $i^{+}$ and $i^{*}$ bands. For other bands such as FUV and NUV, wide aperture total fluxes were measured first and then converted to 3\arcsec aperture flux. This consistency in measuring photometry allows one to determine  accurate colours, leading to small uncertainties in the measurement of the spectral energy distribution (SED) fitting for {\it k}-correction, galaxy classification and measuring the stellar mass.  A more detailed description of the photometric catalogue can be found in \citet{Capak:2007,Ilbert:2009}. 

Table~\ref{tab:cosmos_cat} lists the photometric bands that we used in our analysis.  We corrected all magnitudes for the zero-point offset listed in Table~\ref{tab:cosmos_cat} following \citet{Ilbert:2009} and then applied the corrections of the Galactic dust extinction using the dust map of \citet{Schlegel:1998}. {\it k}-corrections were also applied for all magnitudes using values obtained through a $\chi^{2}$ template fitting procedure. 

\subsection{Spectroscopic Data}
Our spectroscopic data is taken from the zCOSMOS \citep{Lilly:2007} survey. The zCOSMOS \citep{Lilly:2007} is a large optical redshift survey undertaken in the COSMOS field using the VIMOS spectrograph mounted on the VLT at the European Southern Observatory (ESO), Chile. 
The main goal of the survey was to trace the large-scale structure of the Universe up to $z \sim$~1 and to characterise galaxy groups and clusters. The zCOSMOS survey consists of two distinct parts, called zCOSMOS-bright and zCOSMOS-deep. The zCOSMOS-bright is a magnitude-limited survey ($I_{AB} <$~22.5 measured in the $I$-band of {\it HST} ACS), targeting $\sim$20,000 galaxies in the redshift range of 0.1~$< z <$~1.2 (see Fig.~\ref{fig:z_diagram}). This survey was undertaken on the entire 1.7~deg$^{2}$ COSMOS field. The zCOSMOS-deep surveys $\sim$10,000 galaxies, colour-selected to be in 1.4~$< z <$~3.0, in the central 1~deg$^{2}$ of the COSMOS field.   

\begin{figure*}
\centering
\includegraphics[width=130mm]{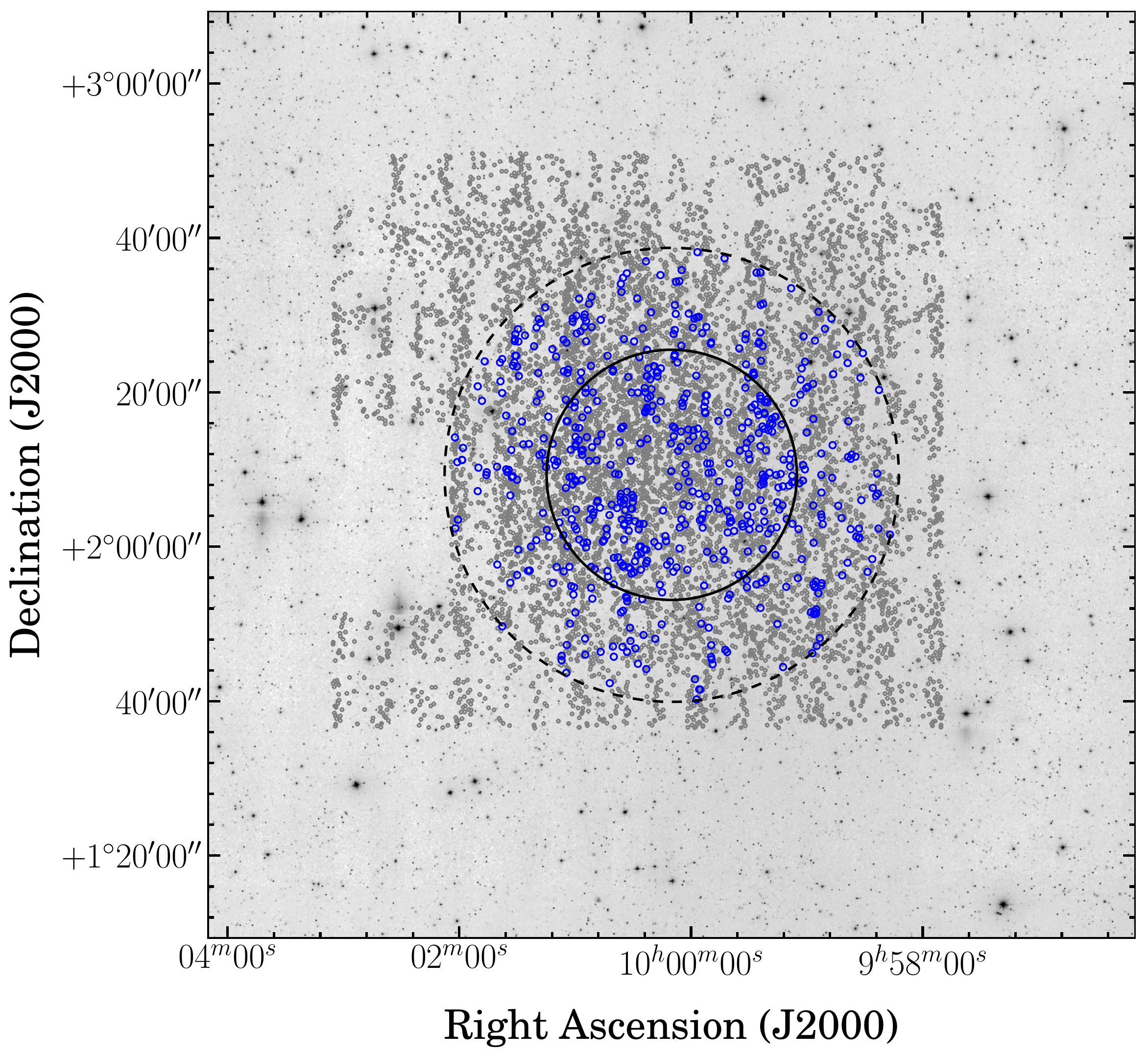} 
\caption[Distribution of objects in the zCOSMOS catalogue]{The spatial distribution of objects in the zCOSMOS catalogue. The blue circles denote galaxies covered by the GMRT beam and frequency. The solid and the dashed circles indicate FWHM (32.25~arcmin) and 10~per~cent level (58.8~arcmin) of the GMRT primary beam at 1040~MHz, respectively.}
\label{fig:object_distribution}
\end{figure*}

In this paper the spectroscopic data used comes from the zCOSMOS-bright catalogue which has spectra for 10,644 objects, the so-called 10k-bright sample. These contain a statistically complete subset of 10,109 objects. Spectra obtained from the zCOSMOS-bright survey cover a wavelength range of approximately 5550 to 9450~\AA, yielding a spectral resolution of $R \sim$~600 sampled at $\sim$2.5~\AA~pixel$^{-1}$. The velocity uncertainty of the zCOSMOS-bright redshifts is $\sim$110~\kms. For more details about data and data reduction, refer to \citet{Lilly:2007,Lilly:2009}. The zCOSMOS 10k-bright catalogue provides redshifts along with confidence classes indicating reliability of its redshift measurements. The confidence classes \citep[see Table~1 in][]{Lilly:2009} vary from class 0 (no redshift obtained) to class 4 (most secure redshift) with additional class 9 for one-line redshifts where the line is believed to be either [OII] or {\Ha}. For the analysis in this paper, we restrict the sample to galaxies with the most secure redshift, i.e. those in class of 3 or 4. This selection produces a sample of 506 redshifts with reliable redshifts that lie within the GMRT data cube ($\sim$1~deg$^{2}$ at the observed frequency 1040 MHz) and the redshift range of 0.35~$< z <$~0.39 (see also Fig.~\ref{fig:object_distribution}).   

\begin{figure*}
\centering
\includegraphics[width=84mm]{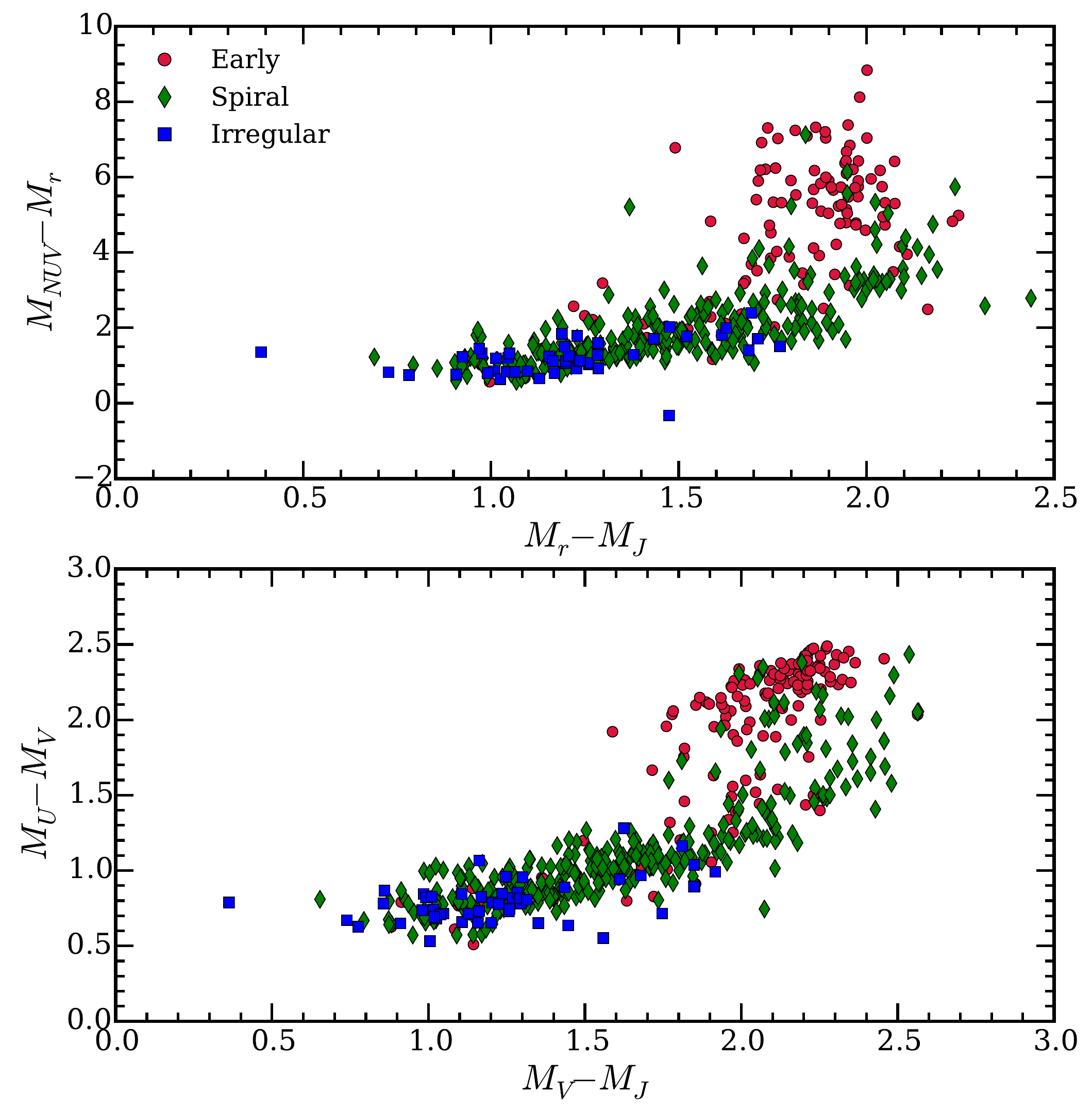}
\includegraphics[width=84mm]{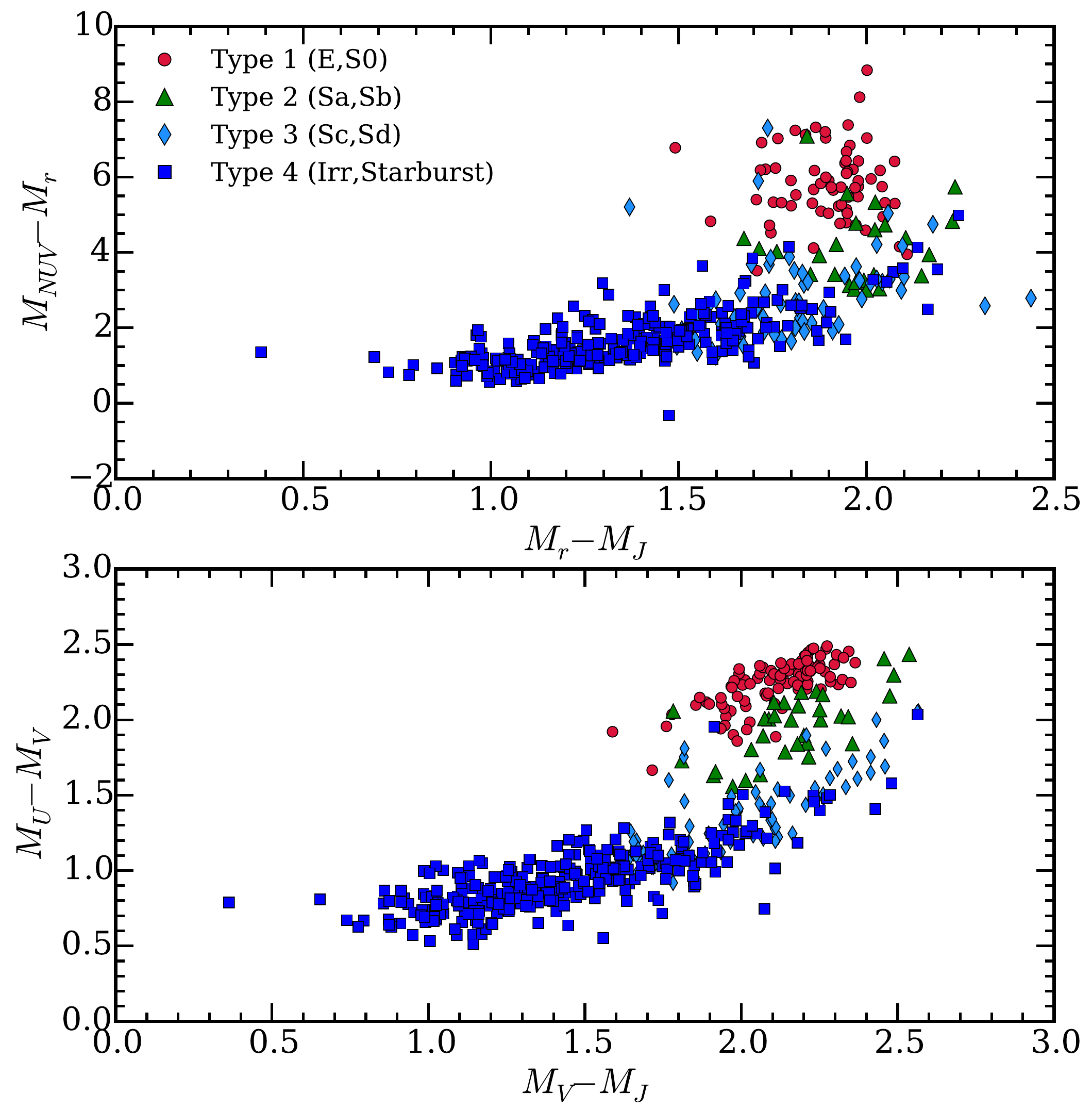} 
\caption[colour--colour diagrams of subsamples classified by morphological and spectro-photometric approaches]{colour--colour diagrams of subsamples classified by morphological ({\it left} panels) and spectro-photometric ({\it right} panels) approaches. The upper panels are $NUV-r_+$ and $r_{+}-J$ colour--colour diagrams and the lower panels are $u^{*}-V$ and $V-J$ colour--colour diagrams. In each panel on the left, morphological subclasses of early-types, spirals and irregulars are shown in red circles, green diamonds and blue squares, respectively. Four subsamples of the spectro-photometric approach from Type~1 to Type~4 are denoted on the right panels by red circles, green triangles, blue diamonds and squares.}
\label{fig:ccd_diagram}
\end{figure*}

\subsection{Galaxy Classification}

We used two different methods to classify our sample galaxies: one based on galaxy morphology and the other on spectro-photometry. The COSMOS field has the {\it HST} ACS imaging data \citep{Koekemoer:2007} with sufficient depth and resolution to perform morphological analysis. The COSMOS archival database provides a morphological class catalogue, which is based on applying an automatic and objective morphological classification technique to high-quality {\it HST} images \citep{Cassata:2007,Tasca:2009}. The morphological classification divides the galaxies into  three morphological classes, i.e., early-types including ellipticals and lenticulars, spirals, and irregulars.

\begin{figure}
\centering
\includegraphics[width=86mm]{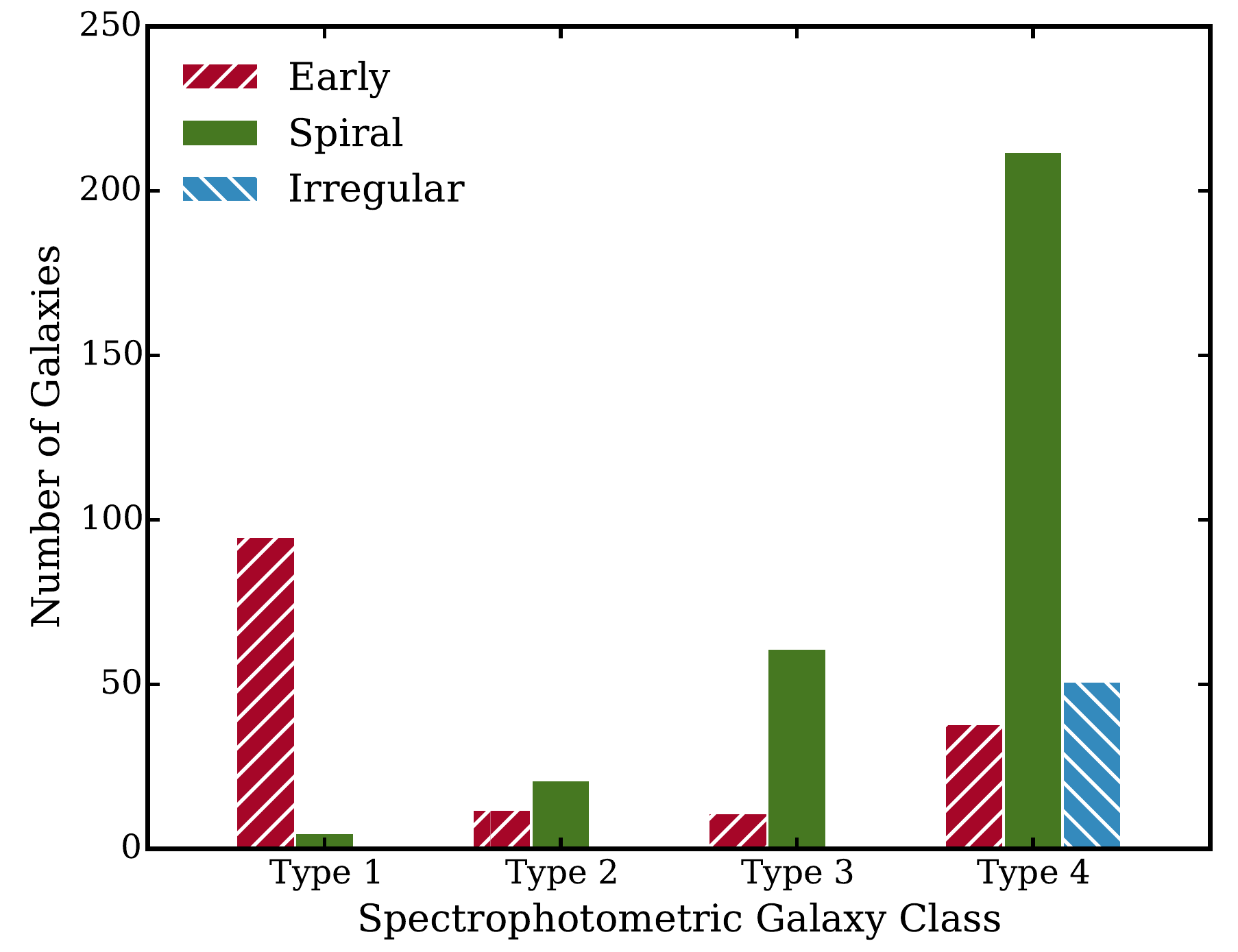} 
\caption{The spectro-photometric classes of galaxies morphologically classified. Red ($\slash$), green, and blue ($\backslash$) bars denote morphological classes--early, spiral, and irregular, respectively.}
\label{fig:comp_morph_phot}
\end{figure}

The spectro-photometric classification is based on matching the rest-frame magnitudes and colours to a set of templates. Here, {\sc le phare}\footnote{\url {http://www.cfht.hawaii.edu/~arnouts/lephare.html}} \citep{Arnouts:1999,Ilbert:2006}, a $\chi^{2}$ template fitting code, using 20 photometric magnitudes from the COSMOS photometry catalogue, was applied in combination with spectroscopic redshifts from the zCOSMOS catalogue for each galaxy in the sample. This spectro-photometric classification divides our sample galaxies into four types; early-type E/S0 (Type~1), early spirals Sa/Sb (Type~2), late spirals Sc/Sd (Type~3), and irregular and starburst galaxies (Type~4). 

\begin{figure}
\centering
\includegraphics[width=86mm]{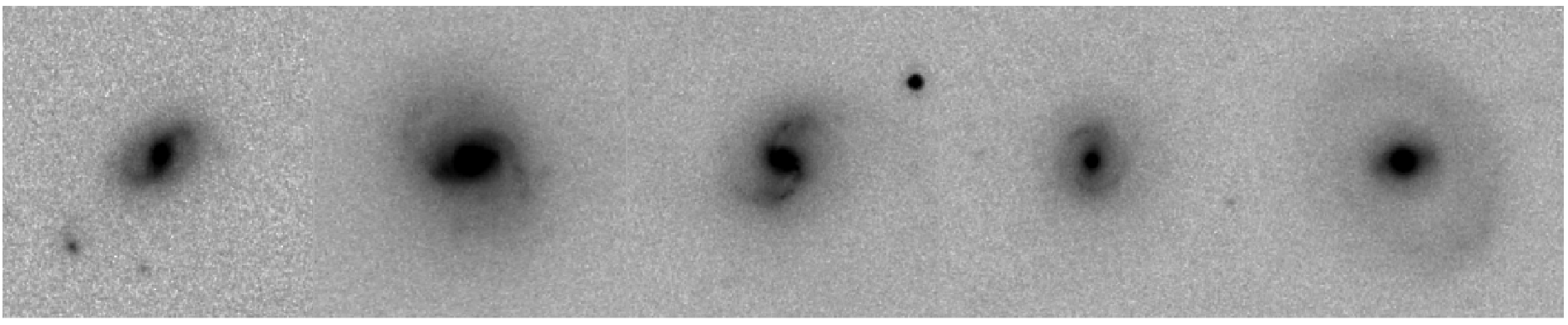} 
\includegraphics[width=86mm]{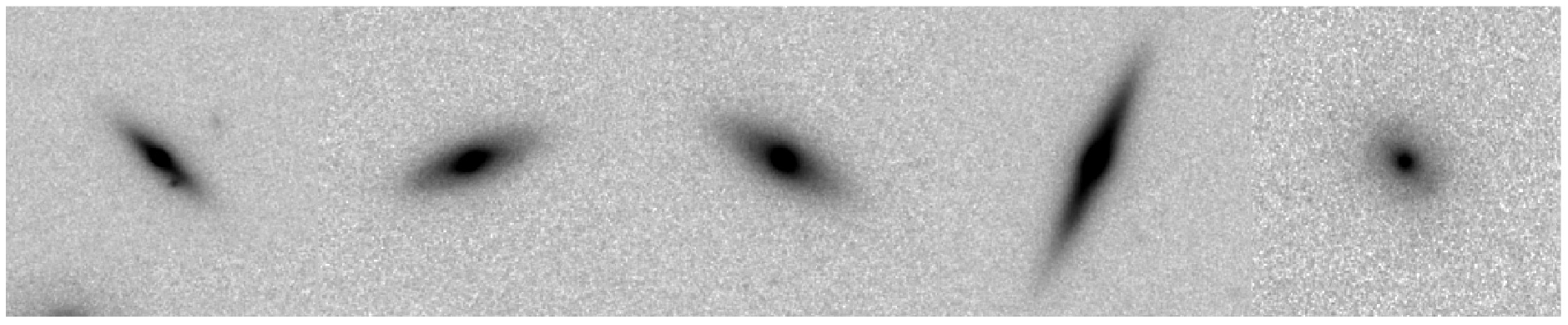} 
\caption{{\it HST} cutout images of blue early-type ({\it upper row}) and red spiral ({\it lower row}) examples. The size of each thumbnail image is 6\arcsec by 6 \arcsec. The images were obtained using G10/COSMOS image cutout tool.}
\label{fig:HST_cutout}
\end{figure}

Following \citet{Williams:2009,Ilbert:2009} we used $u^{*}-V$ vs. $V-J$ and $NUV-r_+$ vs. $r_{+}-J$ colour--colour diagrams to compare these two classification schemes. In Fig.~\ref{fig:ccd_diagram}, the left-hand panels plot the morphological classes of the galaxies, while the right-hand panels show the spectro-photometric classes. As can be seen, the morphologically classified early-types and spirals are not sharply separated in these diagrams. In particular, there are galaxies which are morphologically classified as early type but are blue in colour. Fig.~\ref{fig:ccd_diagram} shows that spectro-photometry provides a much cleaner separation. Fig.~\ref{fig:comp_morph_phot} summarises the leakage of the morphological classes between the spectro-photometric classes.

Visual inspection of the {\it HST} images of the blue early types shows that some of them have an obvious spiral or ring feature as seen in the upper row of Fig.~\ref{fig:HST_cutout}. This indicates that the uncertainty of the morphological classification can affect the {\HI} stacking analysis for different galaxy types that we do below. Although the spectro-photometry clearly separates the early types, there exist Type~1 galaxies having spiral morphology (called ``red spirals''). These galaxies look highly-inclined or edge-on spirals in their {\it HST} images (see the lower row of Fig.~\ref{fig:HST_cutout}). Their high inclination can cause more internal reddening by dust, resulting in redder colour and mis-matched template. These galaxies have a negligible effect on the {\HI} stacking analysis because the fraction of such galaxies is below 5~\%. We have adopted the spectro-photometric classification in our further analysis.

\section{GMRT Radio Data}

\subsection{Observation}

The zCOSMOS field was observed for a total of 134~hours using the GMRT. The observations were conducted over 20 days spread over the years 2008 and 2009. The total observation time includes 115~hours of on-source time with the remaining time spent on calibrator scans. The central frequency of the GMRT observations was 1040~MHz and the bandwidth was 32 MHz corresponding to an {\HI} redshift range of 0.345~$< z <$~0.387. The observations were done using the old hardware correlator which divided the 32~MHz bandwidth into two 16~MHz-wide sidebands. Each sideband had two polarisations and 128 spectral channels, giving a channel width of 0.125~MHz ($\sim$36.3 \kms at $z =$ 0.37). The pointing centre of the GMRT observations was R.A.~10$^{\rm h}$00$^{\rm m}$10$\rm \fs$01~Dec.~+02$^{\rm d}$19$^{\rm m}$19$\rm \fs$95 (J2000). The primary beam size (FWHM) of the GMRT is approximately 32.25~arcmin at 1040~MHz and the 10$\%$ beam level is 58.8~arcmin which was the limit used for selecting galaxies for the stacking analysis. Observations of 3C~48, 3C~147 and 3C~286 were used to calibrate the flux density scale. 0943-083 served as a phase calibrator.

\begin{figure}
\centering
\includegraphics[width=84mm]{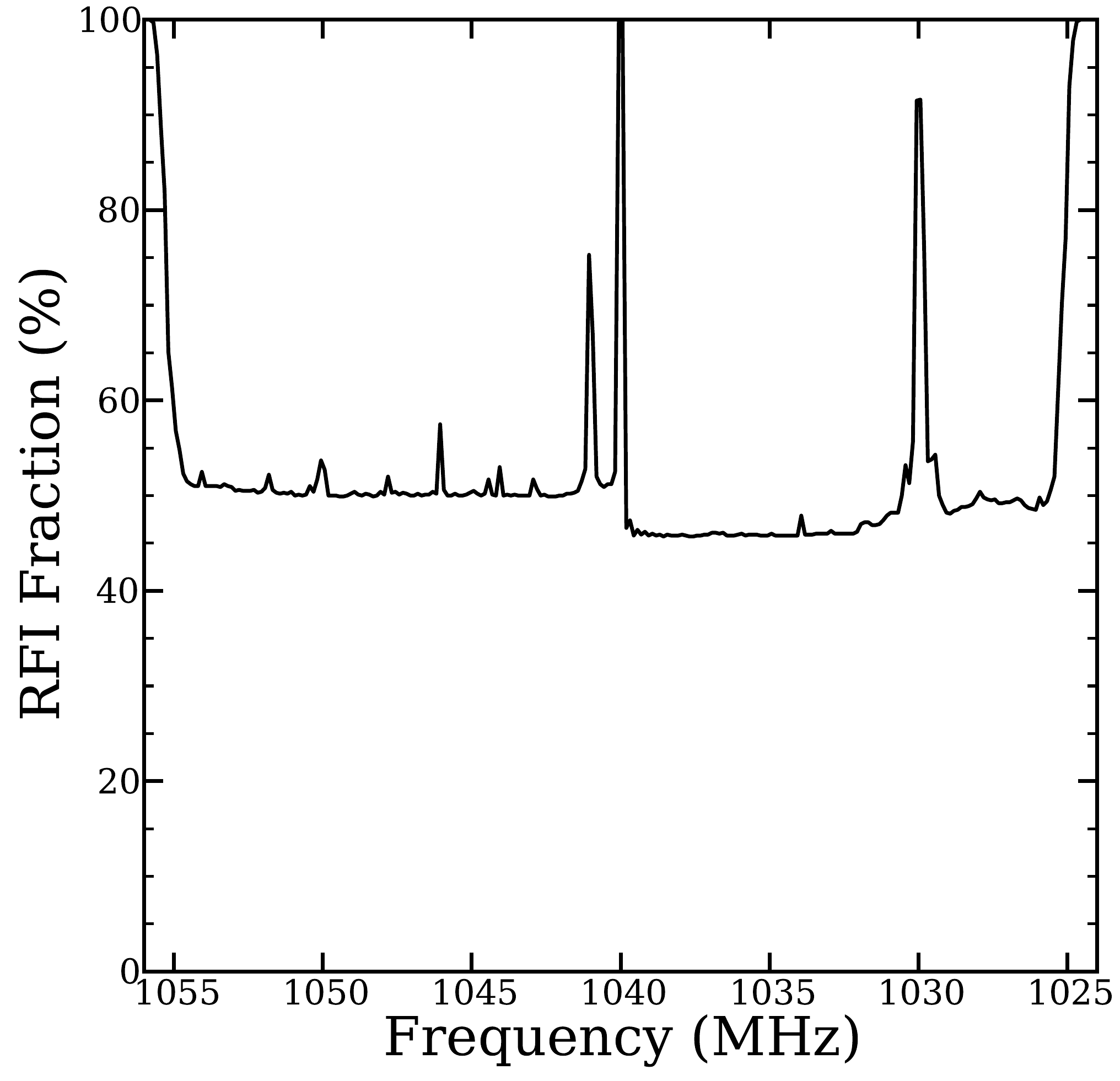} 
\caption[Fraction of flagged data]{The fraction of flagged data after all data reduction procedures.}
\label{fig:rfi_frac}
\end{figure}

\begin{figure*}
\centering
\includegraphics[width=120mm]{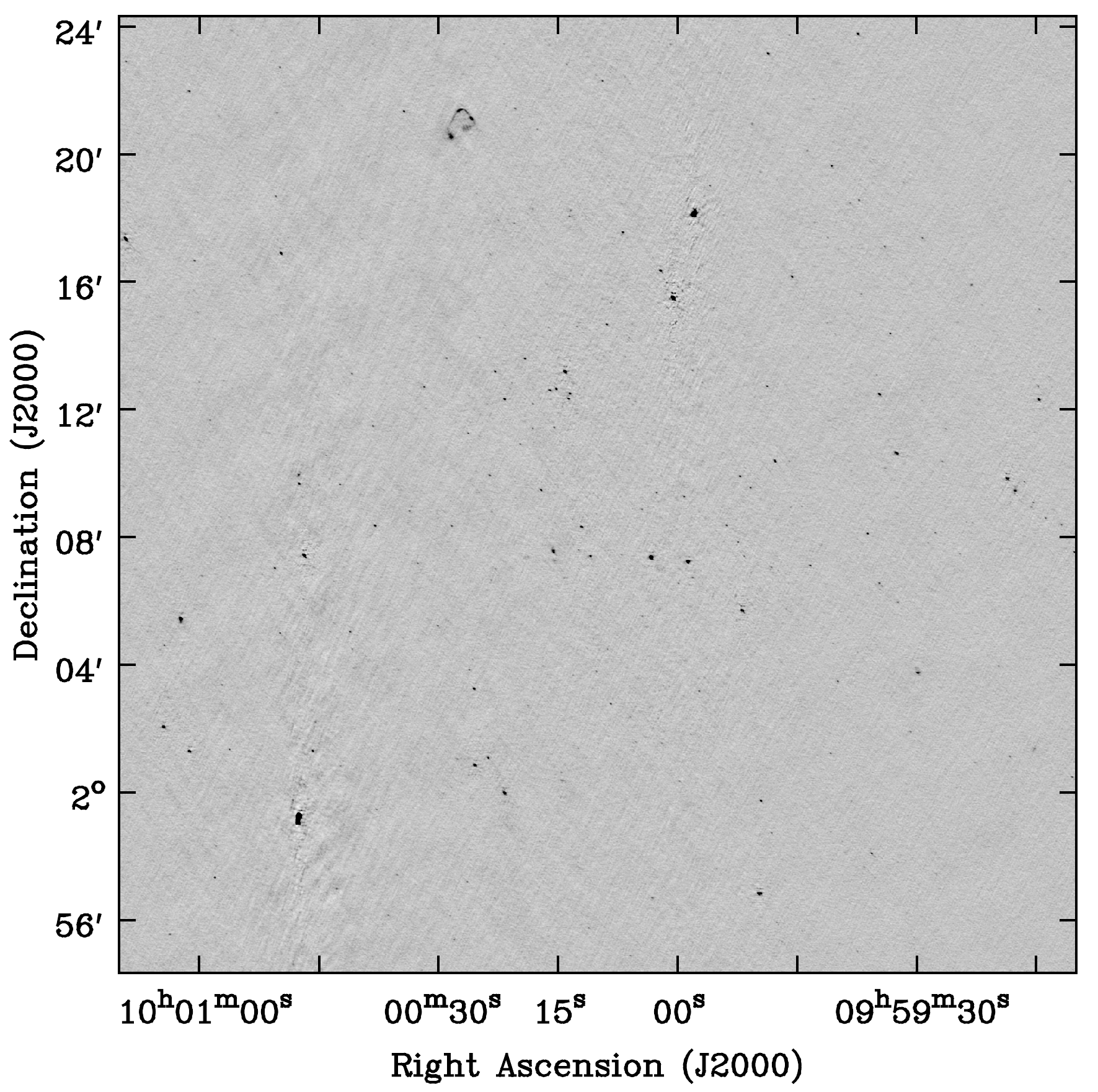} 
\caption[The GMRT continuum image of the COSMOS field]{The central 30$\arcmin$ area of the entire GMRT continuum image of the COSMOS field.}
\label{fig:contmap_zcosmos}
\end{figure*}

\subsection{Data Reduction}

The GMRT data reduction of the COSMOS field followed a standard reduction procedure including flagging, calibration and imaging. The GMRT data were first processed using {\sc flagcal} \citep{Prasad:2012, Chengalur:2013}, an automated flagging and calibration software developed for the GMRT data. As seen in Fig.~\ref{fig:rfi_frac}, the frequency range of the COSMOS field covered by the GMRT was affected by radio frequency interference (RFI); there were also several instrumental malfunctions during the observations that caused several antennas to be unusable. The edges of each sideband were completely flagged, which lead to the peak around 1040~MHz in Fig.~\ref{fig:rfi_frac} where 100 per~cent of the data has been flagged. 
Data near 1030~MHz appear to be severely contaminated by RFI. It is known that aircrafts often generate RFI at this frequency. In addition, the GMRT front-end system had a limited dynamic range during the time when these observations were done, which can generate intermodulation products from strong RFI bursts. This system has since been significantly improved. About half the data in each sideband  has been flagged during iterations
through the data reduction process. 

Subsequent processing was done using the Common Astronomy Software Applications ({\sc casa})\footnote{{\url {http://casa.nrao.edu}}} package. If necessary, additional flagging was carried out manually, using {\sc casa} plotting and editting tools. Broad band flux and phase calibration was done using the central 80 channels, and the bandpass was determined using scans of both the flux and phase calibrators. After the calibration solutions were determined and applied, the COSMOS field datasets for each day were split and re-inspected separately as continuum images to check whether there remained any bad data appearing as artefacts in the images. Data for the 20 observing runs were then concatenated separately for each sideband. By design \citep{Scoville:2007} the COSMOS field does not contain any bright radio sources.  We found that self-calibration did not significantly improve the quality of the images, and hence the final analysis was done using images without any self-calibration.

\subsection{Continuum Image}

To make the final continuum image of the COSMOS field at $z \sim$~0.37, the central 100 channels from each sideband (lower and upper sideband) were selected and concatenated. In order to avoid bandwidth smearing, each sideband used 100 channels subdivided into 10 channel averages during the imaging. A $\sim$1$\times$1~deg$^2$ continuum image was made with pixel size 0.9\arcsec~pixel$^{-1}$,  ``robust'' weighting \citep{D_Briggs:1995} with a robust value of 0, and the $w$-projection \citep{Cornwell:2008} algorithm. The rms noise in the central regions of the final continuum (Fig.~\ref{fig:contmap_zcosmos})  is $\sim$12.3 $\mu$Jy~beam$^{-1}$ with a synthesised beam resolution of 3.5\arcsec $\times$ 2.4\arcsec. The astrometric accuracy of the GMRT imaging was determined by comparing the positions of sources detected in this image with high signal-to-noise ratio ( $\geq$ 5 $\sigma$) against the positions listed in the VLA FIRST survey catalogue \citep{Becker:1995}. The average positional offset found was $\sim$0.68 arcsec which is less than the image pixel size of 0.9 arcsec. The maximum measured astrometric offset was 1.8 arcsec, which is still significantly smaller than the GMRT synthesised beam.

\subsection{Line Data Cube}

The spectral data cubes for the two sidebands were made with the same pixel size, robust weighting and wide-field imaging algorithm as were used to make the continuum image. The synthesised beam of the data cubes is 3.5\arcsec $\times$ 2.4\arcsec, corresponding to  $\sim$17.9~$\times$~12.0 kpc$^{2}$ at $z \sim$~0.37. The final spectral data cubes of each sideband were made by subtracting the continuum from these data cubes. For continuum subtraction, clean components of individual continuum sources were subtracted from the $uv$ data using the {\sc casa} task `\texttt{uvsub}'and any residual continuum flux was removed in the `$image$' domain by the {\sc casa} task `\texttt{imcontsub}'. The rms noise levels per frequency channel of each sideband are $\sim$140~$\mu$Jy~beam$^{-1}$ (lower sideband) and $\sim$117~$\mu$Jy~beam$^{-1}$ (upper sideband), respectively.

\section{Stacked {\HI} Emission and {\HI} Mass Measurements}

\begin{figure}
\centering
\includegraphics[width=84mm]{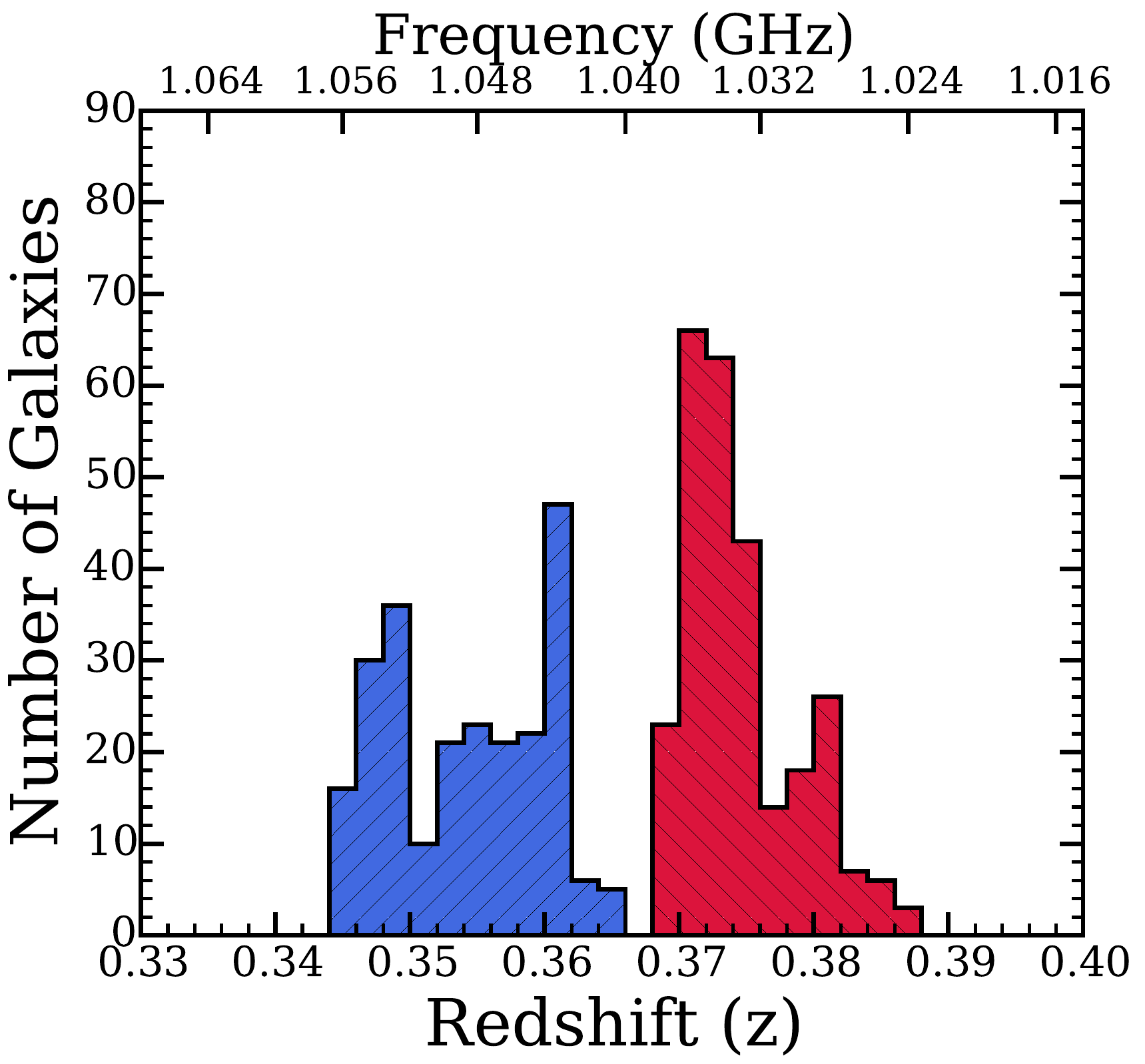} 
\caption[The redshift distribution of COSMOS sample observed by the GMRT]{The redshift distribution of the COSMOS sample observed by the GMRT. Blue and red hatch areas are galaxies that lie in the upper and lower sidebands, respectively.}
\label{fig:z_hist}
\end{figure}

\begin{figure*}
\centering
\includegraphics[width=120mm]{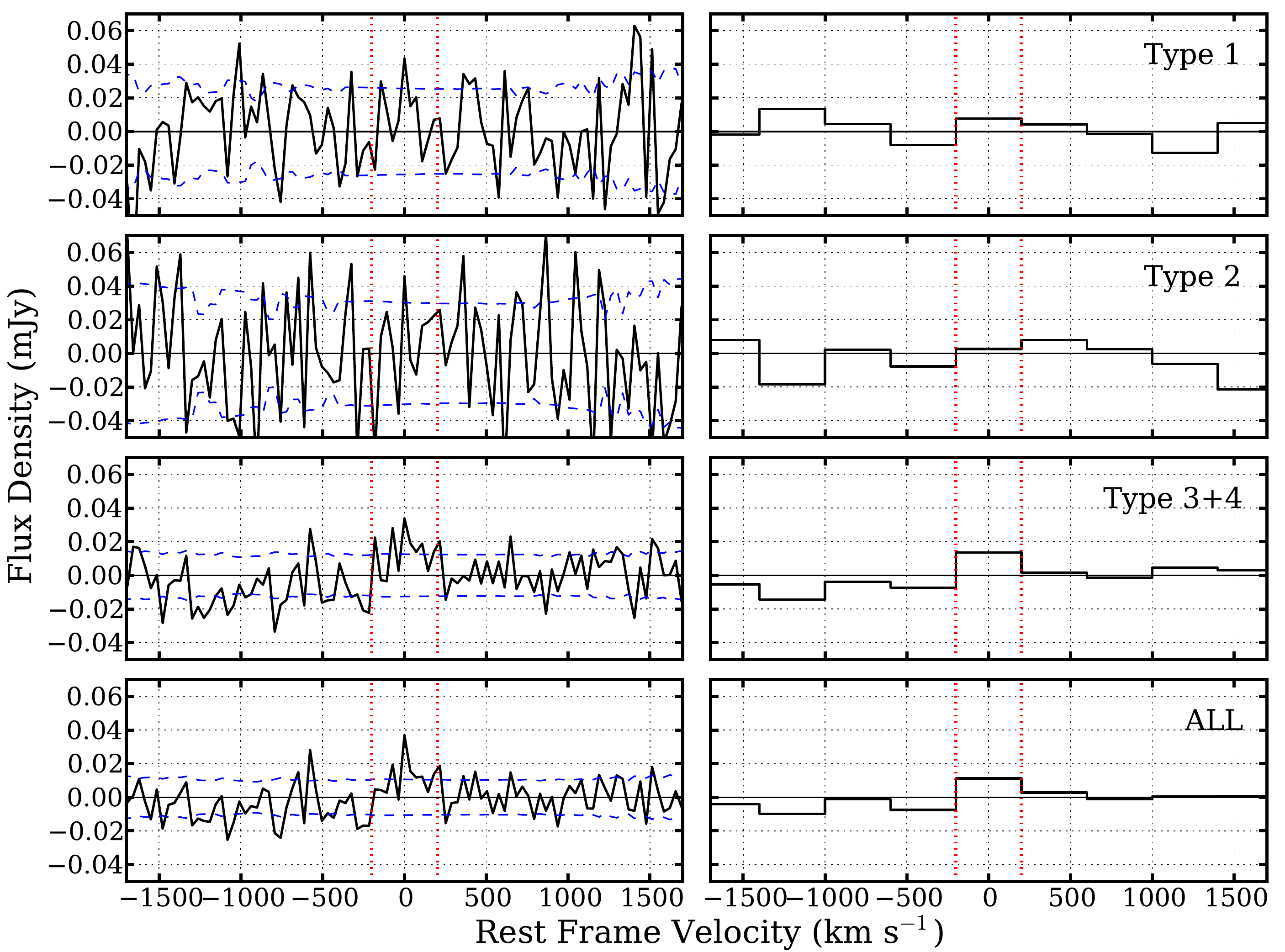} 
\caption[Stacked {\HI} spectra for each subsample]{Stacked {\HI} spectra of galaxies in each subgroup ({\it left} panels). The {\it right} panels show the re-binned stacked spectra with a velocity width of 500~\kms (the vertical dotted lines). The horizontal dashed line in each {\it left} panel is the 1$\sigma$ error of the stacked spectrum.}
\label{fig:stacked_zcosmos}
\end{figure*}

Before performing the {\HI} stacking analysis, an automated source-finding software \citep[{\sc duchamp},][]{Whiting:2012} was used to search for any directly detected {\HI} emitters in the final data cubes. An eyeball inspection was also done. No significant source was found by either method. 

As plotted in Fig.~\ref{fig:z_hist}, the lower and upper sideband data cubes contain 269 and 237 sample galaxies, respectively. However, some of the galaxies lie in channels seriously affected by RFI or at the edges of the data cubes. These galaxies were excluded, leaving 474 galaxies for the {\HI} stacking analysis. The usable 474 galaxies consist of 95 galaxies with spectro-photometric Type~1, 58 with Type~2 and 321 with Type 3 or 4. Using their known positions and redshifts, the spectra for these galaxies were extracted from the data cubes with dimensions of 1$\degr \times$1$\degr \times$128 channels and corrected for the primary beam attenuation. The GMRT primary beam pattern was assumed to have a circular Gaussian profile given by 

\begin{equation}
 \label{eq:pbcor}
gain = e^{-(2\sqrt{{\rm ln}2}d/\theta)^{2}}, \quad \theta = 26.2 \times \frac{1280~{\rm MHz}}{f_{\rm obs}},
\end{equation}
where $f_{\rm obs}$ is the observing frequency, $\theta$ is the Half Power Beam Width (HPBW) and $d$ is angular separation from the GMRT pointing centre, given in units of arcmin. The HPBW was taken from the measurements provided in the National Centre for Radio Astrophysics (NCRA) website\footnote{{\url {http://www.ncra.tifr.res.in}}}. After this correction the spectra  were shifted and aligned to the same rest frame velocity. The stacked spectrum was computed from a weighted-average, using the rms noise of each primary beam corrected spectra. Stacked spectra were computed separately for the different galaxy samples, as well as for the full 474 galaxy sample.

Fig.~\ref{fig:stacked_zcosmos} shows the {\HI} stacked spectra for each of the subtypes of galaxies. Most of the {\HI} gas in the COSMOS field observed by the GMRT resides in the Type~3 or 4 (late spiral, irregular and star burst galaxies). As expected there is no statistically significant signal from the  early-type galaxies.

To calculate the average {\HI} mass from the co-added spectra for each subsample, the following equation \citep{Wieringa:1992} was used:

\begin{equation}
 \label{eq:H_mass}
  \frac{{\MHI}}{\msun} = \frac{236}{(1+z)} \left( \frac{D_L}{{\rm Mpc}} \right)^2 
  \bigg( \frac{\int S_V dV}{{\rm mJy \,km~s^{-1}}} \bigg),
\end{equation} 
where $z$ is redshift, $D_{L}$ is the luminosity distance in units of Mpc, and $\int S_V dV$ is the integrated {\HI} emission flux in units of $\rm {mJy~\kms}$. The redshift of 0.37 is the median redshift value of stacked galaxies for this calculation, which is also used for the luminosity distance in Eq.~\ref{eq:H_mass}. In calculating the integrated {\HI} flux from the above equation, we must specify the width of velocity window within which all the {\HI} emission flux is contained. This velocity window is estimated from the Tully-Fisher relation \citep{Tully:1977}. For our late-type sample galaxies, the mean $w_{20}$ is 278.8~\kms and the maximum is $\sim$571~\kms. Taking into account the redshift uncertainty of the zCOSMOS survey, $\pm$110~\kms, a velocity width of 500~\kms was used to calculate the integrated {\HI} flux with the stacked {\HI} spectra. 

The average {\HI} masses for each galaxy type are listed in Table~\ref{tab:HI_measure}. The errors in the average {\HI} masses were estimated by applying a jackknife resampling method \citep{Efron:1982}. As expected, the late type (Type~3 and Type~4) galaxies have higher average {\HI} gas content than the early type galaxies. The {\MHI}$/L_{B}$ ratio of the late type galaxies is similar with the median {\MHI}$/L_{B}$ of the Sc and Sd type galaxies in the local universe \citep{Roberts:1994}.

\begin{figure}
\centering
\includegraphics[width=84mm]{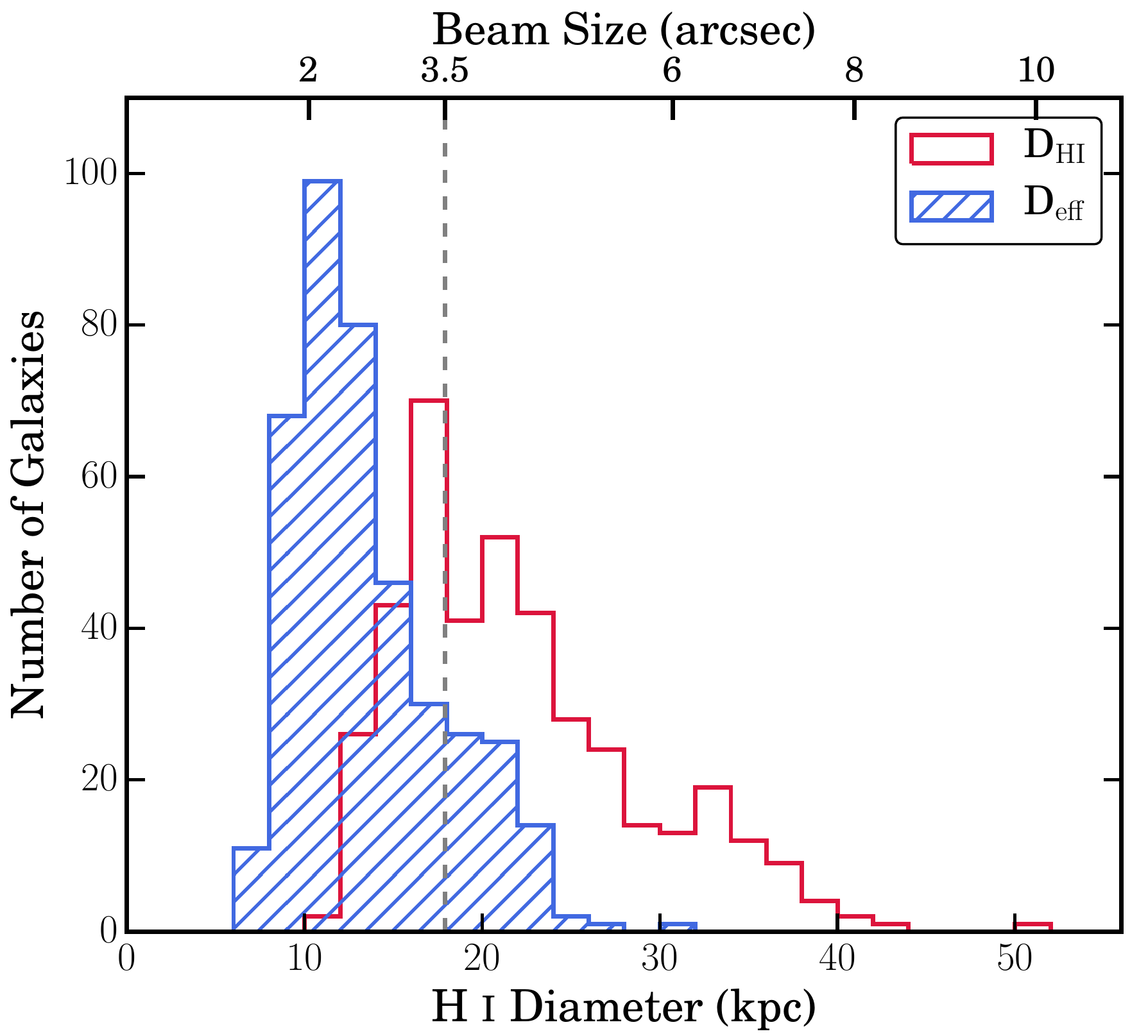} 
\caption[{\HI} size distribution of zCOSMOS sample]{{\HI} size distribution of zCOSMOS sample derived using the relation between optical magnitude and {\HI} size based on \citet{Broeils_Rhee:1997}. Two {\HI} sizes differently defined are used. Blue ($\slash$) and red histograms denote effective {\HI} diameter and {\HI} diameter at a surface density of 1 \msun~pc$^{-2}$, respectively. The GMRT synthesised beam size is shown as the vertical dashed line.}
\label{fig:HI_size_zcosmos}
\end{figure}

The above estimate of the {\HI} mass assumes that the sample galaxies observed in {\HI} are unresolved by the GMRT synthesised beam. To check this assumption the sizes of sample galaxies in {\HI} were estimated from the relationship between optical and {\HI} properties in \citet{Broeils_Rhee:1997}. These authors provided two relationships between the {\HI} diameter and the $B$-band absolute magnitude as follows:
\begin{eqnarray}
  {\rm log}(D_{\HIsub}) & = & (-0.1673\pm0.0142) \times M_{B}-1.9545 \label{eq:relation_opt_HI1}\\ 
  {\rm log}(D_{\rm eff}) & = & (-0.1674\pm0.0152) \times M_{B}-2.1689 \label{eq:relation_opt_HI2}, 
\end{eqnarray}
where $D_{\HIsub}$ is the {\HI} diameter (in kpc) at a surface density of 1 \msun~pc$^{-2}$, and $D_{\rm eff}$ is the diameter containing 50 per~cent of the {\HI} mass. Fig.~\ref{fig:HI_size_zcosmos} shows the distribution of the estimated diameters from the Eq.~\ref{eq:relation_opt_HI1} and \ref{eq:relation_opt_HI2}. The blue and red histograms indicate the distribution of $D_{\rm eff}$ and $D_{\HIsub}$, respectively. As can be seen, the galaxies seem to be partially resolved in terms of $D_{\rm eff}$ by the GMRT synthesised beam of ~3.5$\arcsec$, while $D_{\HIsub}$ of most galaxies is larger than the GMRT synthesised beam. We note however that \citet{Broeils_Rhee:1997} warned that these local correlations might be biased due to the fact that their selection criteria required the galaxies to have large {\HI} mass and optical size. To see whether the GMRT beam partially resolving the large galaxies causes a significant amount of {\HI} flux to be lost, we repeated the analysis using several larger synthesised beams. The data cubes were smoothed to four  additional synthesised beam sizes--5.9$\arcsec$, 7.8$\arcsec$, 9.8$\arcsec$ and 11.7$\arcsec$ which are equivalent to 30, 40, 50, and 60 kpc at $z \sim$~0.37, respectively. The same stacking procedure was conducted for all smoothed data cubes with different beam sizes, and then the average {\HI} masses were calculated and compared. We found no statistically significant change in the stacked signal. For the subsequent calculations in the next sections, we hence assume that there is no significant effect of the GMRT beam size on the {\HI} stacking results.

\begin{table*}
\caption[The measured properties of each galaxy type at $z \sim$~0.37]{The measured properties of each galaxy type at the redshift of $z \sim$~0.37.}
\begin{center}
\label{tab:HI_measure}
 \begin{tabular}{@{}lccccccc}
 \hline
    Sample  & $N_{\rm gal}$ & $\langle \MHI \rangle$ & $\langle L_B \rangle$ & $\langle \MHI \rangle/\langle L_B \rangle$ & $\rho_{L_{B}}$ & $\rhoHI$ & {\it f}  \\
                 &   & (10$^{9}$~{\msun}) & (10$^{9}$~{\lsun}) & ({\msun}/{\lsun}) & (10$^{7}$~{\lsun}~Mpc$^{-3}$) & (10$^{7}$~{\msun}~Mpc$^{-3})$ & \\
 \hline
     Type~1      & 95  & 2.20~$\pm$~2.60 & 17.76~$\pm$~0.04 & 0.12~$\pm$~0.15 & 5.11~$\pm$~0.65 & 0.63~$\pm$~0.75 & - \\
     Type~2      & 58  & 1.50~$\pm$~2.74 & 21.33~$\pm$~0.05 & 0.07~$\pm$~0.13 & 2.69~$\pm$~0.65 & 0.21~$\pm$~0.38 & 1.10~$\pm$~0.09 \\
     Type~3+4 & 321 & 3.83~$\pm$~1.20 & 10.07~$\pm$~0.01 & 0.38~$\pm$~0.12 & 8.12~$\pm$~2.15 & 4.86~$\pm$~1.99 & 1.57~$\pm$~0.02\\
 \hline
     All                & & \multicolumn{5}{c}{{\OHI}$=$~(0.42~$\pm$~0.16)~$\times$~10$^{-3}$} & \\
 \hline
 \end{tabular}
\end{center}
Note: $N_{\rm gal}$ is the number of galaxies that are co-added, $\langle \MHI \rangle$ is the average {\HI} mass per galaxy, $\langle L_B \rangle$ is the mean $B$-band luminosity, $\rho_{L_{B}}$ is the luminosity density and $\rhoHI$ is the {\HI} density that the correction factor in the last column has been applied to. {\it f} is the correction factor for incomplete sampling of the luminosity function.
\end{table*}

\section{Cosmic {\HI} Mass Density ({\OHI})}

The measured {\HI} mass from the stacked {\HI} spectra can be converted to an {\HI} density ({\rhoHI}) and a cosmic {\HI} density ({\OHI}) allowing one to examine how the {\HI} gas content of the Universe evolves over cosmic time. The {\OHI} measurement of the COSMOS field is important because it is the highest-redshift measurement ever made with an {\HI} spectral stacking technique. While there has been a previous measurement of the {\HI} content of galaxies at $z \sim 0.37$ \citep{Lah:2009}, the sample for that study was  galaxies in a rich cluster (Abell~370). Hence unlike the current work, the results from \citet{Lah:2009} cannot be used to study the evolution of the gas content of field galaxies. 

In order to determine {\OHI}, the first step is to derive {\HI} density from the average {\HI} mass ({\MHI}) measured above. Simply dividing {\MHI} by the survey volume does not take into account the incompleteness of our sample to be stacked or the effect of cosmic variance. Following \citet{Rhee:2013} and \citet{Delhaize:2013}, we made use of a volume normalisation method and a correction factor to account for these effects. This approach to derive {\rhoHI} adopts the optical luminosity ($L$) as a weight. We determined the ratio of {\MHI} to $L$ which we then multiplied by the optical luminosity density \citep[e.g., see][]{Fall:1993}. This assumes that all of the {\HI} gas is located in galaxies with optical counterparts \citep{Briggs:1990,Fall:1993,Rao:1993}. Several studies show that this is a reasonable assumption in the local universe. Firstly, the agreement between {\OHI} measurements based on optically selected galaxies and those based on {\HI}-selected samples shows that there is little neutral hydrogen gas associated with galaxies faint or no optical counterparts \citep{Fall:1993,Rao:1993,Zwaan:1997}. Furthermore large blind 21-cm surveys have found that optically invisible but gas-rich galaxies below optical detection threshold do not exist in numbers sufficiently large to bias the inventory of {\rhoHI} \citep{Taylor:2005,Doyle:2005}. 

Accurate photometric measurements of luminosity, which are needed to normalise the stacked {\HI} measurement, are available for the COSMOS field
as discussed in Section 2.2. 
Thanks to a wealth of photometric and spectroscopic data available, the luminosity functions and luminosity densities have been quite well measured up to $z \sim 1$ in the COSMOS field \citep{Zucca:2009}. Since our galaxy classification scheme is the same as that used in \citet{Zucca:2009}, the luminosity function and luminosity density derived in that paper can be directly adopted to derive {\rhoHI}.  The equations below are used to derive {\HI} density for each galaxy type: 

\begin{equation}
  \sum \MHI = \langle \MHI \rangle \times N_{\rm gal}, \qquad {\rhoHI} = \frac{\sum \MHI}{\sum L_B} \times \rho_{L_B}(\it z),
 \label{eq:rhohi_zcosmos}
\end{equation} 
where $\langle \MHI \rangle$ denotes the average {\HI} mass measured using the {\HI} stacking technique, $N_{\rm gal}$ is the number of co-added galaxies, $L_{B}$ and $\rho_{L_{B}}$ are luminosity and luminosity density of the zCOSMOS sample galaxies in the $B$-band, respectively. We separately calculated the {\HI} gas density for each galaxy type of Type~1, Type~2 and Type~3+4. However, this {\HI} density is calculated without accounting for galaxies fainter than the optical survey limit. In the nearby universe, the low-luminosity late-types are known to be gas-rich, so we need to make a first order compensation for the fact that these objects are not all represented in our sample. To correct for this incomplete sampling of the luminosity function, correction factors for each type were obtained using the luminosity function parameters such as the faint-end slope ($\alpha$) and characteristic luminosity ($L^*$) of the luminosity function given by \citet{Zucca:2009}. Refer to Appendix~A in \citet{Rhee:2013} for more details about the correction factor calculation. This correction factor was not applied to early-type galaxies because the contribution of faint early-type population to {\MHI} and {\rhoHI} is small. Moreover the correction factor for the early-type sample (Type~1) made no difference in calculation of the total {\rhoHI}. These correction factors as well as the corrected {\rhoHI} for each type are listed in Table~\ref{tab:HI_measure}. 

\begin{figure*}
 \centering
 \includegraphics[width=150mm]{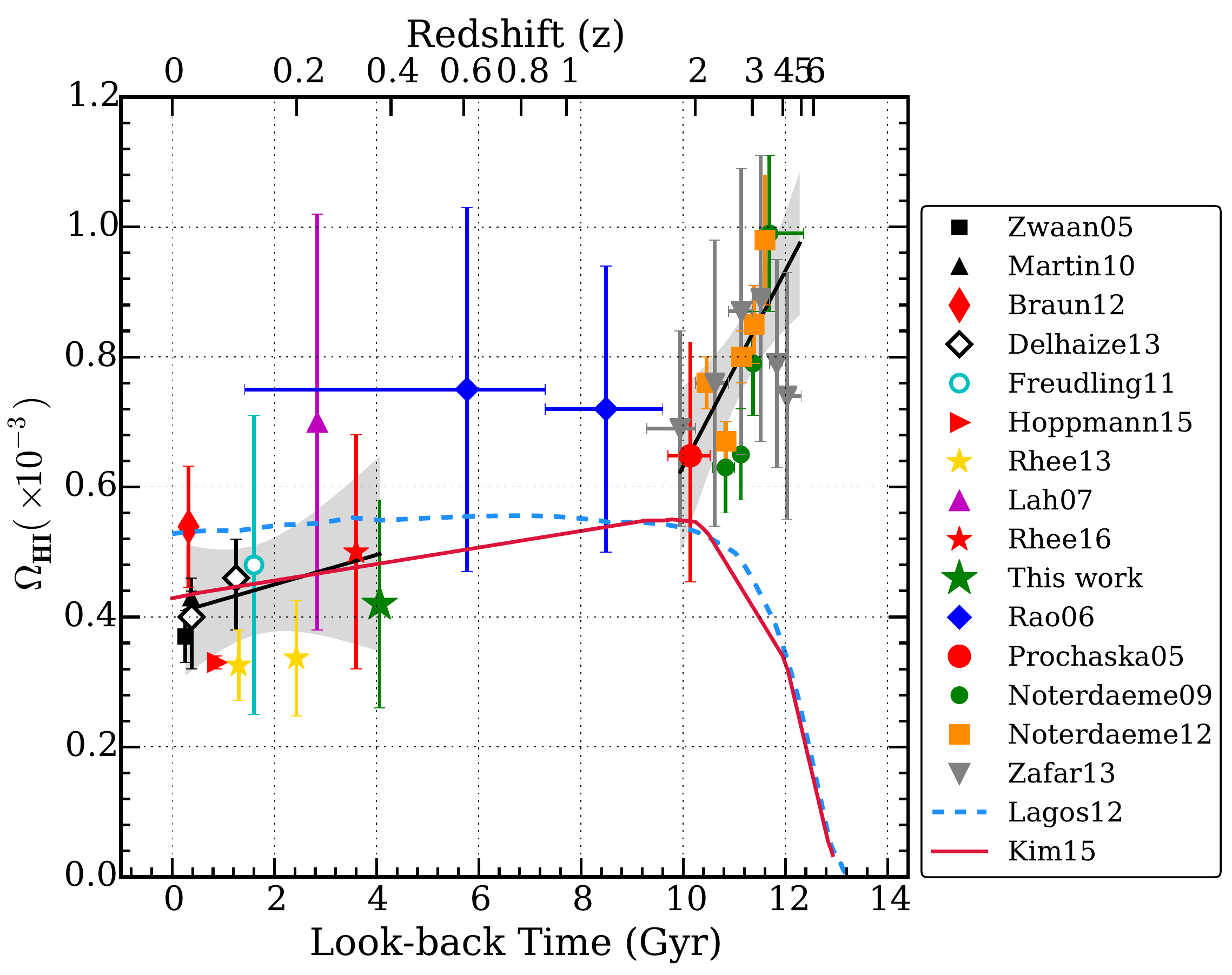}
 \caption[The cosmic {\HI} gas density ({\OHI})  at $z \sim$~0.4]{This shows the cosmic {\HI} gas density ({\OHI}) measurements as a function of redshift ({\it top} axis) and look-back time ({\it bottom} axis). All measurements have been corrected to the same cosmological parameters. Our {\OHI} measurement of the COSMOS field is presented by the green star. The small black square and triangle at $z \sim$~0 are the HIPASS and ALFALFA 21-cm emission measurements by \citet{Zwaan:2005, Martin:2010}, respectively. The red diamond is not a measured value but the average {\OHI} of the two measurements at $z = 0$ to which a correction factor for self-opaque effect has been applied \citep{Braun:2012}. The open diamonds are from the Parkes telescope and an {\HI} stacking technique \citep{Delhaize:2013}. The cyan open circle is the preliminary result from the AUDS \citep{Freudling:2011}. The red right-pointing triangle is the 21-cm direct detection measurement from 60\% data of the total AUDS survey \citep{Hoppmann:2015}. Two yellow stars are measured by \citet{Rhee:2013} using the WSRT and stacking technique. The big purple triangle is measured by \citet{Lah:2007} using the GMRT 21-cm emission stacking. The red star is the {\OHI} measurement of VVDS~14h field at $z \sim$~0.32 \citep{Rhee:2016} using the GMRT along with stacking technique. The blue diamonds, red big circle, green circles, and orange squares are damped Lyman-$\alpha$ measurements from the {\it HST} and the SDSS by \citet{Rao:2006}, \citet{Prochaska:2005}, \citet{Noterdaeme:2009}, \citet{Noterdaeme:2012}, respectively. The grey downward triangles at high redshift of $z >$~2 are ESO UVES measurements of DLAs and sub-DLAs by \citet{Zafar:2013}. The black lines with grey shade areas are least-square fits and their 95~per~cent confidence interval with all {\OHI} measurements at lower redshifts and higher redshifts, respectively. The blue dashed and the red solid lines are model predictions of {\OHI} taken from \citet{Lagos:2014} and \citet{Kim:2015}, respectively.}
 \label{fig:omega_HI_zcosmos}
\end{figure*}

Consistent with earlier works, we define the cosmic {\HI} gas density ({\OHI}) as the ratio of {\rhoHI} to the critical density ($\rho_{\rm crit}$): 
\begin{equation}
  {\OHI} = \frac{\rhoHI}{\rho_{\rm crit}} = \frac{8\pi G \rhoHI}{3H_{0}^{2}},
 \label{eq:ohi_zcosmos}
\end{equation} 
where $H_{0}$ is the Hubble constant and $G$ is the gravitational constant. The critical density at present is $\rho_{\rm crit}$ = 2.78~$\times$~10$^{11} h^2$~\msun~Mpc$^{-3}$, where $h=H_{0}/$100 \kms ($h=$~0.7). Since in principle all galaxy types contribute to the cosmic {\HI} density, the total {\rhoHI} is obtained by summing {\rhoHI} contributions from all types. This gives {\OHI}~$=$~(0.42~$\pm$~0.16)~$\times$~10$^{-3}$. This $2.6\sigma$ measurement is shown in Fig.~\ref{fig:omega_HI_zcosmos} along with other available measurements taken from the literature (see the caption for more details). As can be seen, our measurement taken in conjunction with earlier measurements at lower redshifts indicates that there has been no significant evolution in {\OHI} from $z = 0$ to $z \sim 0.4$. The weighted mean average of {\OHI} from all 21-cm measurements at redshifts $z < 0.4$ gives $\OHI = (0.35 \pm 0.01) \times 10^{-3}$.

Regarding the cosmic variance, it is known that the zCOSMOS field suffers less than 10 per~cent cosmic variance in the redshift interval of $z =$~0 to 0.5 \citep{Driver:2010}. The zCOSMOS field surveyed for this paper by the GMRT has a limited redshift interval of 0.345~$< z <$~0.387 and a small sky area compared to the original, which would result in increased cosmic variance. However, since in our {\OHI} calculation above we do a volume normalisation using luminosity density derived from the full sample of the zCOSMOS field, the cosmic variance that we are subject to is the same as that computed for the entire zCOSMOS field. Moreover, we compared the luminosity density of the zCOSMOS field at $z \sim$~0.37 used for volume normalisation to that derived from a large and complete spectroscopic survey such as Galaxy and Mass Assembly \citep[GAMA,][]{Loveday:2012}. The two luminosity densities are in excellent agreement for the overall galaxy population as well as for each galaxy population. This implies that the volume which we used for normalisation can represent the average of the universe at the redshift that we explored.  

The stacking analysis using 21-cm emission carried out in this paper is based on the assumption that {\HI} gas in our galaxies is optically thin. This means that any possible influence of {\HI} self-absorption, which will lead to an underestimation of the {\HI} mass, is negligible. Since it is very difficult to assess the effect quantitatively and statistically, even large blind {\HI} surveys have made only a rough estimate for the effect, i.e., an underestimation of less than 15 per~cent in {\OHI} \citep{Zwaan:1997, Zwaan:2005}. However based on high-resolution maps of the {\HI} distribution in M31, M32 and LMC, \citet{Braun:2012} recently suggested that galaxies contain a significant population of {\HI} clouds with a size of 100~pc and high {\HI} column density ($>$~10$^{23}$~cm$^{-2}$) which are optically thick. He derived a global opacity correction factor of 1.34$\pm$0.05 and applied this to the local {\OHI} measurements resulting in $\sim$34 per~cent increased {\OHI} as seen in Fig~\ref{fig:omega_HI_zcosmos}. Although the sample used to derive the correction factor covers a fairly large range of {\HI} mass the total number of galaxies in his sample is very small. As such, the validity of applying this correction factor to high redshift measurement like ours seems to be unclear. In any case, were this correction to be uniformly applied to all the {\HI} emission surveys, it would result only in a shift of all the values upwards, and would not affect the conclusion that there appears to be no evolutionary trend in {\OHI} at least out to $z \la 0.4$. 

\section{{\HI} Gas Evolution over last 4~Gyr}

In Fig.~\ref{fig:omega_HI_zcosmos}, our {\OHI} measurement of the COSMOS field is compared with other {\OHI} values available in the published literature \citep{Zwaan:2005,Martin:2010,Freudling:2011,Delhaize:2013,Hoppmann:2015,Rhee:2013,Lah:2007,Rhee:2016,Rao:2006,Prochaska:2005,Noterdaeme:2009,Noterdaeme:2012,Zafar:2013}. Two main observational techniques have been used to measure {\OHI}: 21-cm emission observations at low redshifts and damped Lyman alpha absorption (DLA) observations at high redshifts. All measurements using {\HI} 21-cm emission from both direct detection and stacking are in good agreement. At high redshifts ($z >$~2) all {\OHI} measurements from DLA observations are consistent with one another, showing increase in {\OHI} with redshift. Also there seems to be at least 2 times more {\HI} gas amount than at lower redshifts. We note that the {\HI} 21-cm measurements correspond to the {\HI} gas inside galaxies, while the DLA observations measure the total gas, regardless of whether it lies inside or outside galaxies. In principle these could be different quantities, although, as noted above, in the local universe at least, there is no evidence for a large reservoir of {\HI} that lies in optically dark galaxies.

Many galaxy evolution models have recently attempted to predict {\HI} gas densities across cosmic time to reproduce observations \citep[e.g.,][]{Power:2010,Lagos:2011,Duffy:2012,Dave:2013,Lagos:2014,Rahmati:2015,Kim:2015}. However, there is a tension between observations and theoretical models. We show in Fig.~\ref{fig:omega_HI_zcosmos} theoretical predictions for the evolution of {\OHI} from the recent semi-analytic `Lagos12' and `Kim15' models of \citet{Lagos:2014} and \citet{Kim:2015} for comparison between observations and theories. As can be seen, the theoretical models do match the low redshift data, but not the high-redshift DLA based measurements. 
In contrast, a hydrodynamical simulation study of the distribution of {\HI} around high-redshift galaxies at $z >$~ 1 \citep[e.g.,][]{Rahmati:2015} shows that their {\OHI} predictions are in good agreement with the {\OHI} evolution at high redshifts while they disagree with lower-redshift {\OHI}. In some simulations this problem is interpreted as reflecting the difference between what the {\HI} 21cm measurements and the DLA measurements are sensitive to \citep{Altay:2011,Faucher-Giguere:2011,Fumagalli:2011}. The forthcoming surveys by the SKA pathfinders and the SKA itself will be critically important to resolve this issue.

\section{Summary and Conclusion}

We present the results of an {\HI} spectral stacking analysis using GMRT observations of the COSMOS field. Our sample is chosen from the zCOSMOS-bright 10k catalogue \citep{Lilly:2009}. The individual {\HI} 21-cm line spectra obtained from the GMRT are stacked using the known optical positions and redshifts of the galaxies. The {\HI} spectra are separately stacked for galaxy types classified by SED template fitting and then converted to the average {\HI} mass per galaxy. We find that the {\HI} signal comes primarily from late type galaxies, as expected. The inferred {\MHI}$/L_{B}$ ratio is consistent with that of galaxies in the local ($z=0$) universe. Using the average {\HI} mass along with the integral optical $B$-band luminosity of the sample galaxies and the luminosity density of the COSMOS field, a volume normalisation is applied to obtain the cosmic {\HI} density ({\OHI}). We measure {\OHI}~$=$~(0.42~$\pm$~0.16)~$\times$~10$^{-3}$ at $z \sim$~0.37. This $2.6 \sigma$ measurement is the highest-redshift measurement of {\OHI} ever made using {\HI} spectral stacking. The value of {\OHI} that we measure is consistent within the error bars with both the {\OHI} at $z=0$ as measured from large blind 21-cm surveys as well as that measured from other {\HI} stacking experiments. All the {\HI} 21-cm emission measurements to date show no evidence for evolution of {\HI} gas abundance over the last 4 Gyr; the weighted mean of {\OHI} from all 21-cm measurements at $z < 0.4$ is  (0.35~$\pm$~0.01)~$\times$~10$^{-3}$. This value of {\OHI} is however smaller than that measured at $z \ga 2$ from DLA observations. The next generation of radio telescopes will be sensitive enough to detect the {\HI} signal to redshifts greater than $z =$~1 and will be crucial in understanding the evolution of {\OHI} in the redshift range intermediate between $z \sim 0.4$ and the redshifts probed by DLA observations.

\section*{Acknowledgments}
We are grateful to an anonymous referee for helpful comments and suggestions that improved this work. JR would like to thank Lister Staveley-Smith for useful comments. We thank Claudia del P. Lagos and Han-Seek Kim for providing their model predictions used in Fig.~\ref{fig:omega_HI_zcosmos}. We also thank the staff of the GMRT for their assistance. The GMRT is operated by the National Centre for Radio Astrophysics of the Tata Institute of Fundamental Research. This research was funded by an Australian Indian Strategic Research Fund (AISRF) grant. This fund was jointly administered by the Department of Innovation, Industry, Science and Research in Australia and by the Department of Science and Technology in India. The project title was ``Gas in Galaxies in the Distant Past''. Parts of this research were conducted by the Australian Research Council Centre of Excellence for All-sky Astrophysics (CAASTRO), through project number CE110001020. The G10/COSMOS cutout tool uses data acquired as part of the Cosmic Evolution Survey (COSMOS) project and spectra from observations made with ESO Telescopes at the La Silla or Paranal Observatories under programme ID 175.A-0839. The G10/COSMOS cutout tool is hosted and maintained by funding from the Inertnational Centre for Radio Astronomy Research (ICRAR) at the University of Western Australia. Full details of the catalogue can be found in \citep{Davies:2015} or on the G10 website: {\url {http://ict.icrar.org/cutout/G10/}}.

\bibliographystyle{mn2e}
\bibliography{zCOSMOS_reference}

\label{lastpage}
\end{document}